\documentclass[11pt,twoside]{article}
\usepackage{graphicx}
\usepackage{amsmath}
\usepackage{amssymb}
\usepackage{mathrsfs}

\begin{document}

\thispagestyle{plain}

\newcommand{\be}{\begin{equation}}
\newcommand{\ee}{\end{equation}}
\newcommand{\ds}{\displaystyle}
\newcommand{\bdm}{\begin{displaymath}}
\newcommand{\edm}{\end{displaymath}}
\newcommand{\bea}{\begin{eqnarray}}
\newcommand{\eea}{\end{eqnarray}}
\newcommand{\mbf}{\mathbf}
\newcommand{\rmd}{\mathrm{d}}
\newcommand{\rme}{\mathrm{e}}
\newcommand{\rmi}{\mathrm{i}}
\newcommand{\bm}{\boldsymbol}
\newcommand{\vpint}{-\!\!\!\!\!\!\int}

\begin{center} {\Large \bf
\begin{tabular}{c}
Gauge-independent Husimi functions \\
of charged quantum particles 
in the electro-magnetic field
\end{tabular}
 } \end{center}


\begin{center} {\bf Ya. A. Korennoy}\end{center}


\begin{center}
{\it P. N. Lebedev Physical Institute of the Russian Academy of Science, \\ 
                       Leninskii prospect 53, Moscow 119991, Russia }
\end{center}

\begin{abstract}\noindent
Gauge-independent Husimi function ($Q-$function) of states of charged quantum 
particles in the electro-magnetic field is introduced using the gauge-independent
Stratonovich-Wigner function, 
the corresponding dequantizer and quantizer operators transforming the density matrix
of state to the Husimi function and vice versa are found explicitly,
and the evolution equation for such function is derived.
Also own gauge-independent non-Stratonovich Wigner function is suggested and its 
Husimi function is obtained. Dequantizers and quantizers for these Wigner and Husimi functions are given.

\end{abstract}


\section{Introduction}

Since the very beginning of the creation of quantum mechanics, the question of its formulation
in terms of the distribution function on the phase space, like the classical kinetic theory, 
has attracted the attention of many scientists, despite the fact that the Heisenberg uncertainty relation 
prohibits the existence  of the joint distribution function of the position and momentum in the quantum case.

A great success in this connection was the introduction of the Wigner function \cite{Wigner32}
and the writing of the dynamic equation for it \cite{Wigner32,Moyal1949}.
For the $N-$dimensional system the Wigner function is introduced as the Weyl symbol \cite{Weyl1927} 
of the  density matrix $\hat\rho(t)$ of the state
\be			\label{WigDef}
W(\mathbf{q},\mathbf{P},t)=\frac{1}{(2\pi\hbar)^N}
\int\left<\left.\mathbf{q}-\frac{\mathbf u}{2}\right|\hat\rho(t)\left|\mathbf{q}+\frac{\mathbf u}{2}\right.\right>
\exp\left(\frac{\rmi}{\hbar}\mathbf{u}\mathbf{P}\right)\rmd ^Nu,
\ee
where $\mbf P$ is the generalized momentum 
corresponding to the generalized momentum operator $\hat{\mbf P}=-\rmi\hbar\partial/\partial{\mbf q}$.
The transformation inverse to (\ref{WigDef}) has the form:
\be			\label{RhoFromW}
\left<\mbf q|\hat\rho(t)|\mbf q'\right>=\int W\left(\frac{\mbf q+\mbf q'}{2},\mbf P,t\right)
\exp\left(\rmi\mbf P\frac{\mbf q-\mbf q'}{\hbar}\right)d^NP.
\ee
Despite the fact that the Wigner function can take negative values, it is successfully applied
in many applications since the 1950s (see, e.g., \cite{Silin1, Silin2, Silin3, Silin4, Landau}) and to the present.
The properties of the Wigner function were considered, e.g., in \cite{OConnellWignerPhysRep}.

At the same time, if we smoothly average the Wigner function on the scales of the hyper-volume
$\hbar^N$ of phase space, then it can be made nonnegative.
This way in the article \cite{Husimi40} and later in \cite{Kano1965} the Husimi function 
(the so called $Q-$function) was introduced  
\be			\label{QDef}
Q(\mathbf{q},\mathbf{P},t)=\frac{1}{(2\pi\hbar)^N}\left<\bm\alpha|\hat\rho(t)|\bm\alpha\right>,
\ee
representing up to the normalization factor the transition probability of a quantum system 
from the state $\hat\rho(t)$ into a coherent state
$|\bm\alpha\rangle$ \cite{GlauberPhysRevLett1963,GlauberPhysRev1963}, where 
$\bm\alpha=(2\hbar)^{-1/2}(\lambda^{1/2}\mbf q+\rmi\lambda^{-1/2}\mbf P)$, 
$\lambda=m\omega$. Up to the inessential phase factor the state $|\bm\alpha\rangle$
is defined in the position representation as follows:
\be			\label{CohStateDef}
\langle\mbf q'|\bm\alpha\rangle=\left(\frac{\lambda}{\pi\hbar}\right)^{N/4}
\exp\left[-\frac{\lambda}{2\hbar}(\mbf q-\mbf q')^2-\frac{\rmi}{\hbar}\mbf P(\mbf q-\mbf q')\right].
\ee
As is known, the state $|\bm\alpha\rangle$ minimizes the Heisenberg uncertainty relation, 
and the remarkable property of the $Q-$function is its connection with the Wigner function 
by means of the formula: 
\be			\label{WignerToHusimi}
Q(\mathbf{q},\mathbf{P},t)=\frac{1}{(\pi\hbar)^N}\int\exp\left(-\frac{\lambda}{\hbar}(\mbf q-\mbf q')^2
-\frac{1}{\lambda\hbar}(\mbf P-\mbf P')^2\right)W(\mbf q',\mbf P',t)d^Nq'd^NP',
\ee
representing the averaging of the function $W(\mbf q,\mbf P,t)$ 
in the phase space with respect to a Gaussian distribution centered at the point $(\mbf q,\mbf P)$.
Formula (\ref{WignerToHusimi}) can be rewritten in the operator form,
\be			\label{WigToHusOperator}
Q(\mathbf{q},\mathbf{P},t)=\exp\left(\frac{\hbar}{4\lambda}\partial_{\mbf q}^2
+\frac{\lambda\hbar}{4}\partial_{\mbf P}^2\right)W(\mbf q,\mbf P,t),
\ee
and the inverse transform of (\ref{WigToHusOperator}) is obviously has the shape:
\be			\label{HusToWigOperator}
W(\mathbf{q},\mathbf{P},t)=\exp\left(-\frac{\hbar}{4\lambda}\partial_{\mbf q}^2
-\frac{\lambda\hbar}{4}\partial_{\mbf P}^2\right)Q(\mbf q,\mbf P,t).
\ee
Also the inverse transform of (\ref{WignerToHusimi}) can be expressed
as a repeated integral,
\be			\label{HusimiToWigner}
W(\mathbf{q},\mathbf{P},t)=\int\frac{d^Nud^Nv}{(2\pi)^{2N}}\int\exp\left[\frac{\hbar\mbf u^2}{4\lambda}
+\rmi\mbf u(\mbf q-\mbf q')+\frac{\lambda\hbar\mbf v^2}{4} +\rmi\mbf v(\mbf P-\mbf P')\right]Q(\mbf q',\mbf P',t)d^Nq'd^NP'.
\ee 

Expansion of  formula (\ref{WignerToHusimi}) to the classical (non-quantum) case
allows to determine the Husimi function $Q_{\mathrm{cl}}(\mbf q,\mbf P,t)$ of the state 
of the classical system having described by the classical distribution function
$W_{\mathrm{cl}}(\mbf q,\mbf P,t)$ as the overlap with a Gaussian distribution.

Consider the motion of a quantum particle having a spin in the 
electromagnetic field with the vector potential $\mathbf{A}(\mathbf{q},t)$
and the scalar potential $\varphi(\mathbf{q},t)$. 
As it is known, the Hamiltonian of such a system has the form \cite{LandauIII}
\be			\label{Hamiltonian}
\hat H=\frac{1}{2m}\left(\hat{\mathbf P}-\frac{e}{c}\mathbf{A}\right)^2
+e\varphi- \hat{\bm\kappa}  \mathbf{B},
\ee
where $\hat{\mathbf P}=-i\hbar\partial/\partial\mathbf{q}$ is a generalized momentum operator,
$m$ and $e$ are mass and charge of the particle, $\mathbf{B}=\mathrm{rot}\mathbf{A}$ 
is a magnetic field strength, $\hat{\bm\kappa}$ is an operator 
of quantum-mechanical  magnetic moment
\be			\label{Moment}
\hat{\bm\kappa}=\frac{\kappa}{s}\hat{\mathbf{s}},
\ee
where $s$ is a spin of the particle, $\hat{\mathbf{s}}$ is a spin operator,
and $\kappa$ is the value of the intrinsic magnetic moment of the particle.

From the classical electrodynamics it is known that 
potentials of the field are defined only up to the gauge transformation
\cite{LandauII}
\be			\label{eq3}
\mathbf{A} ~~\rightarrow ~~\mathbf{A} +\nabla \chi,
~~~~
\varphi~~\rightarrow ~~\varphi-\frac{1}{c}\frac{\partial\chi}{\partial t},
\ee
where $\chi$ is an arbitrary function of spatial coordinates and time.

Since the electric field intensity $\mathbf{E}$ and the magnetic field strength
$\mathbf{B}$ are defined in terms of the potentials as:
\be			\label{eq4}
\mathbf{E}=-\mathrm{grad}\varphi-\frac{1}{c}\frac{\partial}{\partial t}\mathbf{A},
~~~~
\mathbf{B}=\mathrm{rot}\mathbf{A},
\ee
then the gauge transformation (\ref{eq3}) does not affect the values 
of $\mbf E$ and $\mbf B$.
Therefore the part of Hamiltonian (\ref{Hamiltonian}) responsible for
the interaction of the spin with the magnetic field is independent 
on the gauge transformation, and  we can restrict our considerations only to the case
of $s=0$. The generalization to the non-zero spin particles is straightforward 
(see \cite{Korarticle9,Korarticle14,Korarticle12}).

The requirement of invariance of the Schr\"odinger equation 
under the gauge transformation simultaneously with the gauge-independence
of ``probability density''\, $|\Psi|^2$ leads us to the form of the conversion
of the wave function \cite{LandauIII}:
\be			\label{eq5}
\Psi ~~\rightarrow ~~\exp\left(\frac{\rmi e}{c\hbar}\chi \right)\Psi.
\ee
Accordingly, the conversions of the density matrix of the state of the 
system under the gauge transformation acquires the form:
\be			\label{eq6}
\hat\rho_{\mathrm c}=
\exp\left(\frac{\rmi e}{c\hbar}\chi \right)\hat\rho\,\exp\left(-\frac{\rmi e}{c\hbar}\chi \right).
\ee

In \cite{Stratonovich2} the gauge-independent Wigner function was constructed, and its 
evolution equation was derived in \cite{Serimaa1986}. The gauge-invariance in the 
tomographic probability representation of quantum mechanics was considered in \cite{Korarticle12}
(see review articles \cite{IbortPhysScr,MankoMankoFoundPhys2009} about the 
tomographic probability representation).

The aim of this work is introduction of
gauge-independent Husimi function ($Q-$function) of states of charged quantum 
particles in the electro-magnetic field, and  is derivation of the evolution equation for such function.

\section{\label{Art08Section2}Gauge-independent Husimi function}
For the construction of quantum Husimi representations, 
in which the evolution equation would be gauge-independent, 
we need to introduce gauge-independent quantum Husimi functions.
This can be done with the help of a gauge-independent Wigner function
obtained in \cite{Stratonovich2},
\be
W_\mathrm{g}(\mathbf{q},\mathbf{p},t) 
=\frac{1}{(2\pi\hbar)^{3}}\int
\exp\left(\frac{\rmi }{\hbar}\mathbf{u}\left\{\mathbf{p}
+\frac{e}{c}\int_{-1/2}^{1/2} \rmd\tau \mathbf{A}(\mathbf{q}+\tau\mathbf{u},\,t)\right\}\right)
\rho\left(\mathbf{q}-\frac{\mathbf u}{2},\,\mathbf{q}+\frac{\mathbf u}{2},\,t\right)
\rmd^3u,
			\label{WigNew}
\ee
where $\mathbf{p}$ is a kinetic  momentum.

The gauge-independent Husimi function $ Q_\mathrm{g}(\mbf q,\mbf p,t)$ should be introduced
using a formula similar to (\ref{WignerToHusimi})
\be			\label{WignerGToHusimiG}
Q_\mathrm{g}(\mathbf{q},\mathbf{p},t)=\frac{1}{(\pi\hbar)^N}\int\exp\left(-\frac{\lambda}{\hbar}(\mbf q-\mbf q')^2
-\frac{1}{\lambda\hbar}(\mbf p-\mbf p')^2\right)W_\mathrm{g}(\mbf q',\mbf p',t)d^Nq'd^Np'.
\ee
With this definition the formulas (\ref{WigToHusOperator},\,\ref{HusToWigOperator},\,\ref{HusimiToWigner})
of the relations between $Q_\mathrm{g}(\mathbf{q},\mathbf{p},t)$ and $W_\mathrm{g}(\mbf q',\mbf p',t)$ 
remain valid if we replace the generalized momentum $\mbf P$ by the kinetic momentum $\mbf p$ in them.

Combining formulas (\ref{WigNew}) and (\ref{WignerGToHusimiG}) we can write
\be			\label{QgFromRho}
Q_\mathrm{g}(\mathbf{q},\mathbf{p},t)=\int \langle\mbf q_1|\hat\rho(t)|\mbf q_2\rangle 
\langle\mbf q_2|\hat U_{Q_\mathrm{g}}(\mbf q,\mbf p)|\mbf q_1\rangle d^3q_1d^3q_2,
\ee
where we introduce the matrix element for the corresponding dequantizer operator
\bea
\langle\mbf q_2|\hat U_{Q_\mathrm{g}}(\mbf q,\mbf p)|\mbf q_1\rangle&=&
\frac{1}{(2\pi\hbar)^3}\left(\frac{\lambda}{\pi\hbar}\right)^{3/2}
\exp\Bigg\{-\frac{\lambda}{2\hbar}(\mbf q-\mbf q_2)^2-\frac{\lambda}{2\hbar}(\mbf q-\mbf q_1)^2
\nonumber \\[3mm]
&+&\frac{\rmi}{\hbar}(\mbf q_2-\mbf q_1)\left[\mbf p
+\frac{e}{c}\int_{-1/2}^{1/2}\mbf A\left(\frac{\mbf q_2+\mbf q_1}{2}+\tau(\mbf q_2-\mbf q_1),t\right)d\tau\right]\Bigg\}.
			\label{matrUQg}
\eea
From (\ref{matrUQg}) we can see that $\hat U_{Q_\mathrm{g}}(\mbf q,\mbf p)$ is Hermitian  operator, 
so the Husimi function $Q_\mathrm{g}(\mathbf{q},\mathbf{p},t)$ is real.

The explicit form of the operator $\hat U_{Q_\mathrm{g}}(\mbf q,\mbf p)$ can be written in the integral form
\be			\label{UQgint}
\hat U_{Q_\mathrm{g}}(\mbf q,\mbf p)=\int\frac{d^3ud^3v}{(2\pi\hbar)^6}
\exp\left[-\frac{\lambda\mbf u^2}{4\hbar}-\frac{\mbf v^2}{4\hbar\lambda}
-\frac{\rmi}{\hbar}(\mbf u\mbf p+\mbf v\mbf q)\right]
\exp\left\{\frac{\rmi}{\hbar}\left[\mbf u\left(\hat{\mbf P}-\frac{e}{c}\mbf A(\hat{\mbf q},t)\right)
+\mbf v\hat{\mbf q}\right]\right\}.
\ee 

The transformation inverse to (\ref{QgFromRho}) can be expressed using the matrix element of 
quantizer operator $\hat D_{Q_\mathrm{g}}(\mbf q,\mbf p)$
\be			\label{RhoFromQg}
\langle\mbf q_1|\hat\rho(t)|\mbf q_2\rangle
=\int \langle\mbf q_1|\hat D_{Q_\mathrm{g}}(\mbf q,\mbf p)|\mbf q_2\rangle
Q_\mathrm{g}(\mathbf{q},\mathbf{p},t) d^3qd^3p,
\ee
where
\bea
\langle\mbf q_1|\hat D_{Q_\mathrm{g}}(\mbf q,\mbf p)|\mbf q_2\rangle&=&
\int\frac{d^3v}{(2\pi\hbar)^3}\exp\Bigg[\frac{\lambda(\mbf q_2-\mbf q_1)^2}{4\hbar}
+\frac{\mbf v^2}{4\hbar\lambda}-\frac{\rmi\mbf v}{2\hbar}(2\mbf q-\mbf q_1-\mbf q_2)
-\frac{\rmi\mbf p}{\hbar}(\mbf q_2-\mbf q_1) \nonumber \\[3mm]
&-&\frac{\rmi e}{\hbar c}(\mbf q_2-\mbf q_1)\int_{-1/2}^{1/2}d\tau
\mbf A\left(\frac{\mbf q_2+\mbf q_1}{2}+\tau(\mbf q_2-\mbf q_1),t\right)
\Bigg].
			\label{matrDg}
\eea
In these formulas it is assumed that $Q_\mathrm{g}(\mathbf{q},\mathbf{p},t)\in S(\mathbb{R}^6)$,
where $S(\mathbb{R}^6)$ is a Schwartz space, and first we take the integral over $d^3q$,
and after that the integrals over $d^3p$ and $d^3v$ are taken.

Assuming a special order of integration, the explicit form of the operator 
$\hat D_{Q_\mathrm{g}}(\mbf q,\mbf p)$ can be written as:
\be			\label{DQgint}
\hat D_{Q_\mathrm{g}}(\mbf q,\mbf p)=\int\frac{d^3ud^3v}{(2\pi\hbar)^3}
\exp\left[\frac{\lambda \mbf u^2}{4\hbar}+\frac{\mbf v^2}{4\hbar\lambda}
-\frac{\rmi}{\hbar}(\mbf{up}+\mbf{vq})\right]
\exp\left\{\frac{\rmi}{\hbar}\left[\mbf u\left(\hat{\mbf P}-\frac{e}{c}\mbf A(\hat{\mbf q},t)\right)
+\mbf v\hat{\mbf q}\right]\right\}.
\ee

\section{\label{Art08Section3}Evolution equation for the gauge-independent Husimi function}
To begin with, let us recall the Liouville equation in the electro-magnetic field 
for the classical distribution function.
For the classical ensemble of non-interacting particles with 
mass $m$ and charge $e$ this equation in the phase space has the form:
\be
\left\{\partial_ t+
\frac{\mathbf{p}}{m}\partial_{\mathbf{q}}
+e\left(\mathbf{E}(\mathbf{q},t)
+\frac{1}{mc}[\mathbf{p}\times\mathbf{B}(\mathbf{q},t)]\right)
\partial_{\mathbf{p}}\right\}W_\mathrm{cl}(\mathbf{q},\mathbf{p},t)=0,
			\label{Liouville}
\ee
where $\mathbf{p}$ is a kinetic momentum, $\mathbf{E}(\mathbf{q},t)$ and 
$\mathbf{B}(\mathbf{q},t)$ are electric and magnetic fields, 
defined by formulas (\ref{eq4}), $W_\mathrm{cl}(\mathbf{q},\mathbf{p},t)$
is a distribution function of non-interacting particles.

The distribution function $W_\mathrm{cl}(\mathbf{q},\mathbf{p},t)$ is independent on 
the gauge transformation \cite{LandauII} 
because the Liouville equation (\ref{Liouville}) includes 
only gauge-independent intensities of the electro-magnetic field.

Gauge-independent Moyal equation for the Wigner function $W_\mathrm{g}(\mathbf{q},\mathbf{p},t)$ has the form \cite{Serimaa1986}:
\be			\label{EqWigNew}
\left\{\partial_t+\frac{1}{m}\left(\mathbf{p}+\triangle\tilde{\mathbf p}\right)\partial_{\mathbf{q}}
+ e\left(\tilde{\mathbf{E}} +\frac{1}{mc}
\left[\left(\mathbf{p}+\triangle\tilde{\mathbf p}\right)\times\tilde{\mathbf{B}} \right]\right)
\partial_{\mathbf p}
\right\}W_\mathrm{g}(\mathbf{q},\mathbf{p},t)=0,
\ee
where
\bdm
\triangle\tilde{\mathbf p}=-\frac{e}{c}\frac{\hbar}{\rmi }
\left[
\frac{\partial}{\partial\mathbf{p}} \times \int_{-1/2}^{1/2}
\rmd\tau\,\tau 
\mathbf{B}
\left(
\mathbf{q}+\rmi\hbar\tau\frac{\partial}{\partial\mathbf{p}},\,t
\right)
\right],
\edm
\bdm
\tilde{\mathbf E}=\int_{-1/2}^{1/2}
\rmd\tau\,
\mathbf{E}
\left(
\mathbf{q}+\rmi \hbar\tau\frac{\partial}{\partial\mathbf{p}},\,t
\right),
~~~~
\tilde{\mathbf B}=\int_{-1/2}^{1/2}
\rmd\tau\,
\mathbf{B}
\left(
\mathbf{q}+\rmi \hbar\tau\frac{\partial}{\partial\mathbf{p}},\,t
\right).
\edm
This equation in the classical limit $\hbar\to 0$ is converted into 
Liouville equation (\ref{Liouville}).

Since the relations between the functions $W_\mathrm{g}(\mathbf{q},\mathbf{p},t)$ and
$Q_\mathrm{g}(\mathbf{q},\mathbf{p},t)$ are the same as between $W(\mathbf{q},\mathbf{p},t)$ and
$Q(\mathbf{q},\mathbf{p},t)$, then the correspondence rules between
operators acting on the Wigner function and the Husimi function do not change.
Consequently, we can write:
\be		\label{CorrespRulesWgQg}
\begin{array} {lcl} 
\mbf q\, W_\mathrm{g}(\mathbf{q},\mathbf{p}) &\leftrightarrow &
\left(\mbf q+\frac{\hbar}{2\lambda}\partial_\mbf{q}\right)Q(\mathbf{q},\mathbf{p},t),
\\
\mbf p\, W_\mathrm{g}(\mathbf{q},\mathbf{p}) &\leftrightarrow &
\left(\mbf p+\frac{\hbar\lambda}{2}\partial_\mbf{p}\right)Q(\mathbf{q},\mathbf{p},t),
\\
\partial_{\mbf{q}} W_\mathrm{g}(\mathbf{q},\mathbf{p}) &\leftrightarrow &
\partial_{\mbf{q}}Q(\mathbf{q},\mathbf{p},t),
\\
\partial_{\mbf{p}} W_\mathrm{g}(\mathbf{q},\mathbf{p}) &\leftrightarrow &
\partial_{\mbf{p}}Q(\mathbf{q},\mathbf{p},t).
\end{array}
\ee
With the help of (\ref{CorrespRulesWgQg}) equation (\ref{EqWigNew}) is transformed to
the evolution equation for the  gauge-independent Husimi function $Q_\mathrm{g}(\mathbf{q},\mathbf{p},t)$
\bea
&&\bigg\{\partial_t+\frac{1}{m}\left(\mbf p+\frac{\hbar\lambda}{2}\partial_{\mbf p}
+\big[\triangle\tilde{\mbf{p}}\big]_Q\right)\partial_{\mbf q} \nonumber \\[3mm]
&&~~~~~~+e\left(\big[\tilde{\mbf E}\big]_Q
+\frac{1}{mc}\left[\left(\mbf p+\frac{\hbar\lambda}{2}\partial_{\mbf p}
+\big[\triangle\tilde{\mbf{p}}\big]_Q\right)\times\big[\tilde{\mbf B}\big]_Q\right]\right)\partial_{\mbf p}
\bigg\}Q_\mathrm{g}(\mathbf{q},\mathbf{p},t)=0,
			\label{EqforQg}
\eea
where
\bea
&&\big[\triangle\tilde{\mbf{p}}\big]_Q=-\frac{e}{c}\frac{\hbar}{\rmi}\left[\partial_{\mbf p}\times
\int_{-1/2}^{1/2}\mbf B\left(\mbf q+\frac{\hbar}{2\lambda}\partial_{\mbf{q}}
+\rmi\hbar\tau\partial_{\mbf{p}},t\right)\tau d\tau\right],
\nonumber \\[3mm]
&&\big[\tilde{\mbf E}\big]_Q=\int_{-1/2}^{1/2}\mbf E\left(\mbf q+\frac{\hbar}{2\lambda}\partial_{\mbf{q}}
+\rmi\hbar\tau\partial_{\mbf{p}},t\right)d\tau,
\nonumber \\[3mm]
&&\big[\tilde{\mbf B}\big]_Q=\int_{-1/2}^{1/2}\mbf B\left(\mbf q+\frac{\hbar}{2\lambda}\partial_{\mbf{q}}
+\rmi\hbar\tau\partial_{\mbf{p}},t\right)d\tau.
\eea
As it should be, equation (\ref{EqforQg}) in the classical limit $\hbar\to0$ is converted 
into the Liouville equation (\ref{Liouville}).

\section{\label{Art08Section4}Non-Stratonovich type of gauge-independent \\
Wigner and Husimi functions}

In the previous sections we considered the Wigner and Husimi functions  on the basis of the definition 
of Stratonovich \cite{Stratonovich2}.
However, it is possible to introduce a gauge-independent Wigner function and its corresponding Husimi function
according to (\ref{WignerToHusimi}, \ref{WigToHusOperator}) by other ways.

Let us define the gauge-invariant dequantizer for the new Wigner function 
$\mathfrak{W_\mathrm{g}}(\mbf q,\mbf p)$ as follows:
\bea
\hat U_\mathfrak{W_\mathrm{g}}(\mbf q,\mbf p)&=&\exp\left[\frac{\rmi e}{\hbar c}\hat{\mbf q}
\int_0^1d\tau\mbf A(\tau\hat{\mbf q},t)\right]
\nonumber \\[3mm]
&\times&\int\frac{d^3ud^3v}{(2\pi\hbar)^6}
\exp\left\{\frac{\rmi}{\hbar}\left[\mbf u\left(\hat{\mbf P}-\mbf p\right)+\mbf v(\hat{\mbf q}-\mbf q)\right]\right\}
\exp\left[-\frac{\rmi e}{\hbar c}\hat{\mbf q}
\int_0^1d\tau\mbf A(\tau\hat{\mbf q},t)\right]. 
			\label{newDequantW}
\eea
Since for the Wigner function the dequantizer-quantizer scheme is self-dual, 
then the following equality takes place for the corresponding quantizer:
$\hat D_\mathfrak{W_\mathrm{g}}(\mbf q,\mbf p)=
(2\pi\hbar)^3\hat U_\mathfrak{W_\mathrm{g}}(\mbf q,\mbf p)$.
Calculation of the matrix element of (\ref{newDequantW}) yields
\bea
\langle\mbf q_2|\hat U_\mathfrak{W_\mathrm{g}}(\mbf q,\mbf p)|\mbf q_1\rangle&=&
(2\pi\hbar)^{-3}\delta\left(\mbf q-\frac{\mbf q_2+\mbf q_1}{2}\right) 
\nonumber \\[3mm]
&\times&\exp\left[\frac{\rmi\mbf p}{\hbar}(\mbf q_2-\mbf q_1)
+\frac{\rmi e}{\hbar c}\mbf q_2
\int_0^1d\tau\mbf A(\tau\mbf q_2,t)
-\frac{\rmi e}{\hbar c}\mbf q_1
\int_0^1d\tau\mbf A(\tau\mbf q_1,t)\right].
			\label{matrnewDequantW}
\eea
Taking into account (\ref{WignerToHusimi}), we find the dequantizer for the corresponding
Husimi function $\mathfrak{Q_\mathrm{g}}(\mbf q,\mbf p)$
\bea
\hat U_\mathfrak{Q_\mathrm{g}}(\mbf q,\mbf p)&=&\exp\left[\frac{\rmi e}{\hbar c}\hat{\mbf q}
\int_0^1d\tau\mbf A(\tau\hat{\mbf q},t)\right]
\int\frac{d^3ud^3v}{(2\pi\hbar)^6}
\exp\left[-\frac{\lambda\mbf u^2}{4\hbar}-\frac{\mbf v^2}{4\hbar\lambda}
-\frac{\rmi}{\hbar}(\mbf{up}+\mbf{vq})\right]
\nonumber \\[3mm]
&\times&
\exp\left[\frac{\rmi}{\hbar}(\mbf u\hat{\mbf P}+\mbf v\hat{\mbf q})\right]
\exp\left[-\frac{\rmi e}{\hbar c}\hat{\mbf q}
\int_0^1d\tau\mbf A(\tau\hat{\mbf q},t)\right]. 
			\label{newDequantQ}
\eea
The matrix element (\ref{newDequantQ}) is obviously equal to the following:
\bea
\langle\mbf q_2|\hat U_{\mathfrak{Q}_\mathrm{g}}(\mbf q,\mbf p)|\mbf q_1\rangle&=&
\frac{1}{(2\pi\hbar)^3}\left(\frac{\lambda}{\pi\hbar}\right)^{3/2}
\exp\bigg[-\frac{\lambda}{2\hbar}(\mbf q-\mbf q_2)^2-\frac{\lambda}{2\hbar}(\mbf q-\mbf q_1)^2
\nonumber \\[3mm]
&+&\frac{\rmi}{\hbar}(\mbf q_2-\mbf q_1)\mbf p
+\frac{\rmi e}{\hbar c}\mbf q_2
\int_0^1d\tau\mbf A(\tau\mbf q_2,t)
-\frac{\rmi e}{\hbar c}\mbf q_1
\int_0^1d\tau\mbf A(\tau\mbf q_1,t) \bigg].
			\label{matrnewUQg}
\eea
From  (\ref{matrnewUQg}) it is obvious that up to the normalization factor the  dequantizer 
$\hat U_\mathfrak{Q_\mathrm{g}}(\mbf q,\mbf p)$ is the projector of the considered state  
$\hat\rho(t)$ onto the pure state $|\Psi_{\mbf{q,p}}\rangle$, i.e. 
$\hat U_\mathfrak{Q_\mathrm{g}}(\mbf q,\mbf p)=(2\pi\hbar)^{-3}|\Psi_{\mbf{q,p}}\rangle\langle\Psi_{\mbf{q,p}}|$,
where  we have up to the phase factor in the position representation:
\be
\langle\mbf q'|\Psi_{\mbf{q,p}}\rangle=
\left(\frac{\lambda}{\pi\hbar}\right)^{3/4}
\exp\bigg[-\frac{\lambda}{2\hbar}(\mbf q-\mbf q')^2
-\frac{\rmi}{\hbar}\mbf p(\mbf q-\mbf q')
+\frac{\rmi e}{\hbar c}\mbf q'
\int_0^1d\tau\mbf A(\tau\mbf q',t)
\bigg].
\ee
Calculations of the matrix element of the quantizer for the function $\mathfrak{Q_\mathrm{g}}(\mbf q,\mbf p)$
give rise to the expression
\bea
\langle\mbf q_1|\hat D_{\mathfrak{Q}_\mathrm{g}}(\mbf q,\mbf p)|\mbf q_2\rangle&=&
\int\frac{d^3v}{(2\pi\hbar)^3}\exp\bigg[\frac{\lambda(\mbf q_2-\mbf q_1)^2}{4\hbar}
+\frac{\mbf v^2}{4\hbar\lambda}-\frac{\rmi\mbf v}{2\hbar}(2\mbf q-\mbf q_1-\mbf q_2)
-\frac{\rmi\mbf p}{\hbar}(\mbf q_2-\mbf q_1) \nonumber \\[3mm]
&-&\frac{\rmi e}{\hbar c}\mbf q_2
\int_0^1d\tau\mbf A(\tau\mbf q_2,t)
+\frac{\rmi e}{\hbar c}\mbf q_1
\int_0^1d\tau\mbf A(\tau\mbf q_1,t)
\bigg].
			\label{matrnewDg}
\eea
When using quantizer (\ref{matrnewDg}), it is assumed that
$\mathfrak{Q}_\mathrm{g}(\mathbf{q},\mathbf{p},t)\in S(\mathbb{R}^6)$,
where $S(\mathbb{R}^6)$ is a Schwartz space, and first we take the integral over $d^3q$,
and after that the integrals over $d^3p$ and $d^3v$ are taken.
With the same stipulation, the explicit form of the operator $\hat D_{\mathfrak{Q}_\mathrm{g}}(\mbf q,\mbf p)$
can be written as:
\bea			\label{newDQgint}
\hat D_{\mathfrak{Q}_\mathrm{g}}(\mbf q,\mbf p)&=&\exp\left[\frac{\rmi e}{\hbar c}\hat{\mbf q}
\int_0^1d\tau\mbf A(\tau\hat{\mbf q},t)\right]
\int\frac{d^3ud^3v}{(2\pi\hbar)^3}
\exp\left[\frac{\lambda \mbf u^2}{4\hbar}+\frac{\mbf v^2}{4\hbar\lambda}
-\frac{\rmi}{\hbar}(\mbf{up}+\mbf{vq})\right]
\nonumber \\[3mm]
&\times&\exp\left\{\frac{\rmi}{\hbar}\left[\mbf u\hat{\mbf P}
+\mbf v\hat{\mbf q}\right]\right\}\exp\left[-\frac{\rmi e}{\hbar c}\hat{\mbf q}
\int_0^1d\tau\mbf A(\tau\hat{\mbf q},t)\right].
\eea

Knowing the quantizers and dequantizers for functions
$\mathfrak{W_\mathrm{g}}(\mbf q,\mbf p)$ and $\mathfrak{Q_\mathrm{g}}(\mbf q,\mbf p)$
one can find the evolution equations for them.
Since $\mathfrak{W_\mathrm{g}}(\mbf q,\mbf p)$ and $\mathfrak{Q_\mathrm{g}}(\mbf q,\mbf p)$
do not depend on the gauge, then their evolution equations must also be gauge-independent.

\section{\label{Art08Section5}Conclusion}
In conclusion, I point out the main results of the paper.
The gauge-independent Husimi function ($Q-$function) of states of charged quantum 
particles in the electro-magnetic field was introduced using the gauge-independent
Stratonovich-Wigner function, 
the corresponding dequantizer and quantizer operators transforming the density matrix
of state to the such Husimi function and vice versa were found explicitly,
and the evolution equation for such function was derived.

Also own non-Stratonovich gauge-independent Wigner function  and its 
Husimi function were suggested and their dequantizers and quantizers were obtained.


\end{document}